\documentclass[preprint,12pt]{elsarticle}
\usepackage{graphics}
\usepackage{subfigure}
\usepackage{graphicx}
\usepackage{amsmath}
\usepackage{float}
\usepackage{amssymb}
\usepackage{bm}

\journal{PhysicaA}
\begin{document}
\begin{frontmatter}



\title{Transport of Coupled Self-propelled Particles with Gaussian Colored Noise}

\author{Bing Wang}
{\ead{hnitwb@163.com}}
\author{Wenfei Wu}
\address{School of Mechanics and Optoelectronics Physics, Anhui University of Science and Technology, Huainan, 232001, P.R.China}

\begin{abstract}
The transport of coupled self-propelled particles with colored noise and potential is investigated. Large translational motion noise intensity is good for the transport in $-x$ direction, but large self-correlation time of translational motion noise will inhibit this transport. For proper value of the asymmetry parameter, coupled passive particles move always in $-x$ direction with increasing angle noise intensity, but coupled self-propelled particles appear the transport reverse phenomenon with increasing angle noise intensity. Large length of the spring is good for the directional transport. The average velocity has a maximum and a minimum with increasing spring constant $k$. For passive and very small self-propelled speed coupled particles, large number of particle is good for directional movement, but the effect of coupling will become weak when the self-propelled speed is large.
\end{abstract}

\begin{keyword}
Self-propelled Particles \sep Average Velocity\sep Transport Reverse


\end{keyword}
\end{frontmatter}

\section{\label{label1}Introduction}
Investigation of property of Brownian particles is a key issue for a variety of situations in recent years due to its ubiquitous importance ranging from physicochemical to biological systems. Some biological processes such as ion pumping, neuronal signaling, porous media, and photosynthesis, rely on the transport of ions\cite{Machura2006,Reguera2006,Angelani,Mei,Lindenberg}. H\"{a}nggi \emph{et al}. presented an overview of artificial Brownian motors and explored new applications of artificial Brownian motors\cite{Hanggi}.

Discussion of active matter has attracted widely attention and shown some interesting phenomena. Ranging from bioinspired micro- and nanorobotics and engines to crowd behavior, the applications of the ideas in active matter research spans a multitude of length scales\cite{DOrsogna,Wu,Koumakis,Li,Liu2016,Guidobaldi,Pietzonka,Ao,Liao,Bertand,Gou,Moreno,Liu}. Romanczuk \emph{et al}. given an overview over the theoretical foundations and concepts of active particles systems and discussed developments in the field of statistical physics applied to active particle systems far from equilibrium\cite{Romanczuk}. Pototsky \emph{et al}. considered a colony of point like self-propelled particles without direct interactions that cover a thin liquid layer on a solid support\cite{Pototsky}. Zhang \emph{et al}. investigated collective motion of self-propelled particles with complex noise environments based on the Vicsek model and found the proportion of noise region has an important impact on the collective motion of the system\cite{Zhang}. Shi \emph{et al}. investigated the effects of the noise and quenched disorder, on the dynamics of active particles in two dimensions and found that within the tailored parameter regime, nonergodic superdiffusion and nonergodic subdiffusion occur\cite{Shi}.

Interactions between particles are of primary importance and should be taken into consideration in a lot of systems. Investigation of these systems exhibits new and interesting properties that we could not find in single-particle systems. Csahok \emph{et al}. studied the motion of a chain of elastically coupled particles in an asymmetric potential and found that the collective behavior of the elastically coupled particles under certain conditions leads to an average velocity which is larger than that of a single particle\cite{Csahok}. Denisov \emph{et al}. studied the overdamped, deterministic dynamics of a chain of charged, interacting particles driven by a longitudinal alternating electric field and additionally interacting with a smooth ratchet potential\cite{Denisov}. Kaviani \emph{et al}. investigated current fluctuations in a stochastic system of classical particles with next-nearest-neighbor interaction\cite{Kaviani}.

In this paper, we investigate the transport phenomenon of coupled self-propelled Brownian particles in the presence of potential and colored noise. The paper is organized as follows. In Section \ref{label2}, the basic model is provided. In Section \ref{label3}, the effects of parameters are investigated by means of simulations. In Section \ref{label4}, we get the conclusions.

\section{\label{label2}Basic model and methods}
In the present work, we consider coupled self-propelled Brownian particles with colored noises in the presence of potential. The dynamics of the $i$th particle is described by the following Langevin equations\cite{Chen}
\begin{equation}
\frac{dx_i}{dt}={v_0}\cos\theta_i+\mu(F_{x_i}+G_{x_i})+\xi_{i}(t) \label{xt}
\end{equation}
\begin{equation}
\frac{dy_i}{dt}={v_0}\sin\theta_i+\mu{G_{y_i}}+\xi_{i}(t) \label{yt}
\end{equation}
\begin{equation}
\frac{d\theta_i}{dt}=\zeta_{{i}}({t}) \label{Thetat}
\end{equation}
where $x_i$ and $y_i$ are the position of the particle. $v_0$ is the self-propelled speed and $\mu$ is the mobility, respectively. $\theta$ is the self-propelled angle and denotes the moving direction. $\xi_i$ is the Gaussian colored noise of translational motion.  $\zeta_i$ is the angle Gaussian colored noise. $\xi_i$ and $\zeta_i$ satisfy the following relations
\begin{equation}
\langle\xi_i(t)\rangle=\langle\zeta_i(t)\rangle=0,
\end{equation}
\begin{equation}
\langle\xi_i(t)\xi_j(t')\rangle=\delta_{ij}\frac{Q}{\tau_Q}\exp[-\frac{|t-t'|}{\tau_Q}],
\end{equation}
\begin{equation}
\langle\zeta_i(t)\zeta_j(t')\rangle=\delta_{ij}\frac{D}{\tau_D}\exp[-\frac{|t-t'|}{\tau_D}],
\end{equation}
$\langle\cdots\rangle$ denotes an ensemble average over the distribution of the random forces. $Q$ and $D$ are the noise intensity of the noises, respectively. $\tau_Q$ and $\tau_D$ are the self-correlation time.

The potential $U(x)$ satisfy the following equation,
\begin{equation}
U({x})=\left\{
\begin{array}{rcl}
\frac{U_0}{L_1}(L_1-x), & & {0<x<L_1}\\
\frac{U_0}{L_2}(x-L_1), & & {L_1<x<L}\\
\end{array}
\right.
\end{equation}
where $L=L_1+L_2$ is the period of the potential. $U_0$ is the potential height. $\Delta=L_2-L_1$ is the asymmetry parameter of the potential. The force $F(x)=-\frac{\partial U(x)}{\partial x}$.

$\vec{G}(i)=G_{x_i}\vec{e}_x+G_{y_i}\vec{e}_y$ is the interaction force due to springs between the nearest-neighbour particles,

\begin{equation}
\vec{G}(1)=k(|\vec{r}_2-\vec{r}_1|-a)\cdot \frac{\vec{r}_2-\vec{r}_1}{|\vec{r}_2-\vec{r}_1|},
\end{equation}
\begin{equation}
\vec{G}(i)=k(|\vec{r}_{i+1}-\vec{r}_i|-a)\cdot \frac{\vec{r}_{i+1}-\vec{r}_i}{|\vec{r}_{i+1}-\vec{r}_i|}+k(|\vec{r}_{i}-\vec{r}_{i-1}|-a)\cdot \frac{\vec{r}_{i}-\vec{r}_{i-1}}{|\vec{r}_{i}-\vec{r}_{i-1}|},  1<i<N-1
\end{equation}
\begin{equation}
\vec{G}(N)=k(|\vec{r}_{N}-\vec{r}_{N-1}|-a)\cdot \frac{\vec{r}_N-\vec{r}_{N-1}}{|\vec{r}_{N}-\vec{r}_{N-1}|},
\end{equation}
here $\vec {r}_i$ is the position vector of the $i$th particle. $a$ is the natural length of the springs. $k$ is the spring constant. $N$ is the number of the particle.

A central practical question in the theory of Brownian motors is the over all long time behavior of the particle, and the key quantities of particle transport are the particle velocity $\langle V\rangle$. After the system reaches a stable state, the average velocity is,
\begin{equation}
\langle V\rangle=\lim_{t\to\infty}\frac{\sum\limits_{i=1}^{i=N}\langle{x_i(t)-x_i(t_0)}\rangle}{N\cdot(t-t_0)}
\end{equation}
$x(t_0)$ is the position of particles at time $t_0$.

\section{\label{label3}Results and discussion}
In order to give a simple and clear analysis of the system. Eqs.(\ref{xt}), (\ref{yt}) and (\ref{Thetat}) are integrated using the Euler algorithm. The total integration time was more than $10^5$ and the integration step time $\Delta t=10^{-4}$. The stochastic averages were obtained as ensemble averages over $10^5$ trajectories. With these parameters, the simulation results do not depend on the time step, the integration time, and the number of trajectories.

\begin{figure}
\centering
\includegraphics[height=10cm,width=12cm]{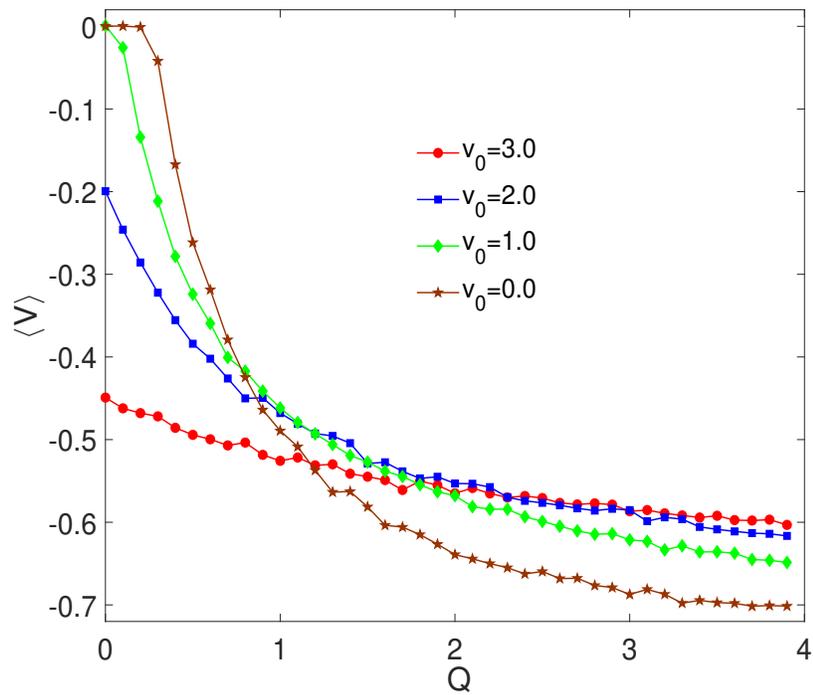}
\caption{The average velocity $\langle V\rangle$ as a function of the translation motion noise intensity $Q$ with different self-propelled speed $v_0$. The particles number $N=4$. The other parameters are $\tau_Q=1.0$, $D=1.0$, $\tau_D=1.0$, $L=2.0$, $\Delta=0.4$, $k=0.5$, $a=1.0$.}
\label{VQ}
\end{figure}

The average velocity $\langle V\rangle$ as a function of the translation motion noise intensity $Q$ with different self-propelled speed $v_0$ is reported in Fig.\ref{VQ}. In this figure, we find the average velocity $\langle V\rangle<0$, this means the coupled particles move in $-x$ direction. The average speed $|\langle V\rangle|$ increases monotonically with increasing $Q$($\langle V\rangle$ decreases monotonically with increasing $Q$). So large $Q$ is good for the directional transport in $-x$ direction. The slope of $\langle V\rangle-Q$ curve is different for different self-propelled speed $v_0$. When $Q$ is small, large $v_0$ is good for directional transport. But when noise intensity $Q$ is large, large $v_0$ will inhabit this directional transport speed. For passive particles($v_0=0.0$), changes of $Q$ has remarkable effect on the average velocity $\langle V\rangle$.
\begin{figure}
\centering
\includegraphics[height=10cm,width=12cm]{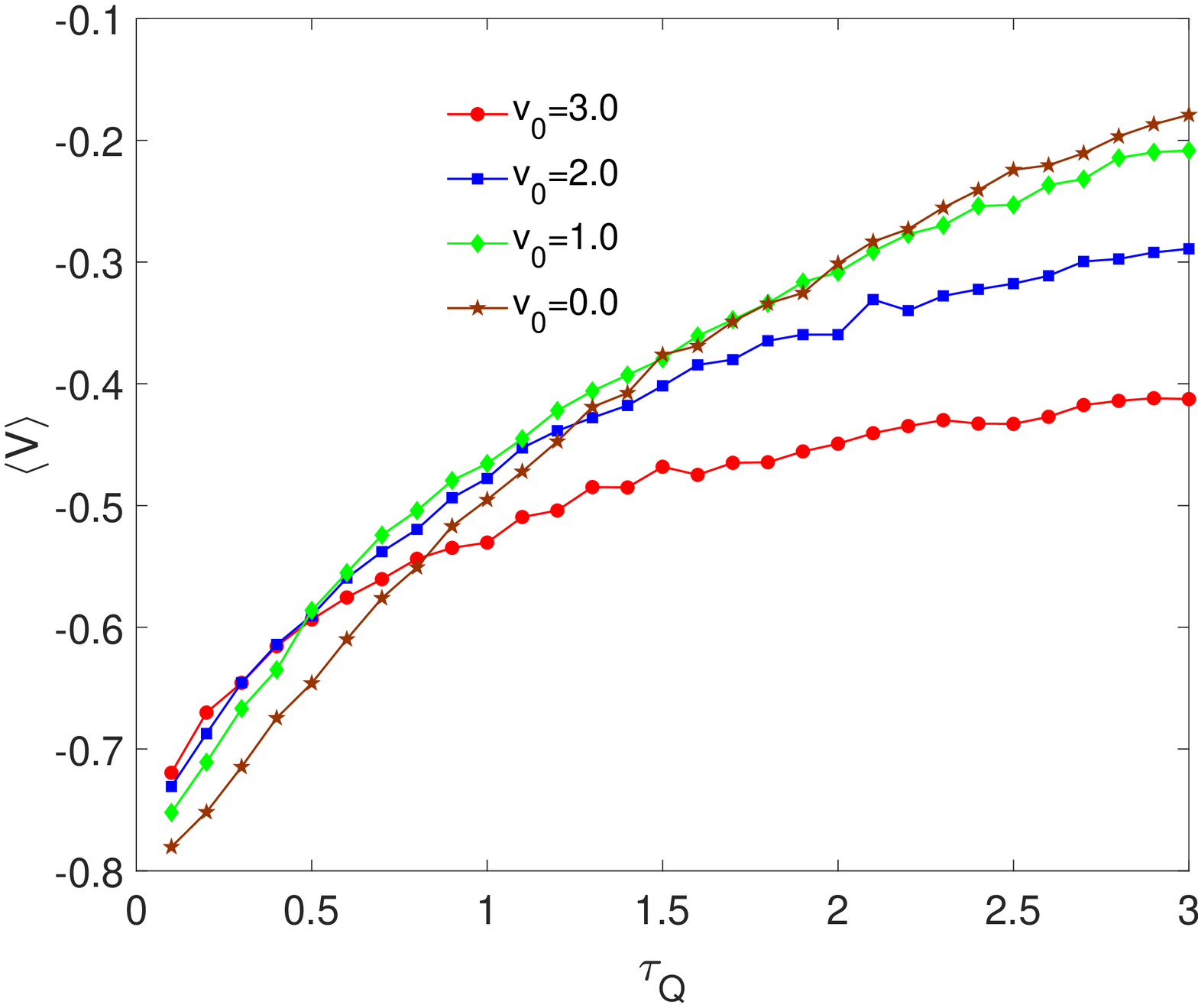}
\caption{ The average velocity $\langle V\rangle$ as a function of the self-correlation time $\tau_Q$ of translation motion noise with different $v_0$. The particles number $N=4$. The other parameters are $Q=1.0$, $D=1.0$, $\tau_D=1.0$, $L=2.0$, $\Delta=0.4$, $k=0.5$, $a=1.0$.}
\label{VtauQ}
\end{figure}

The average velocity $\langle V\rangle$ as a function of the self-correlation time $\tau_Q$ with different $v_0$ is reported in Fig.\ref{VtauQ}.  Contrary to the effect of noise intensity $Q$, we find $\langle V\rangle$ increases with increasing $\tau_Q$(The average speed $|\langle V\rangle|$ decreases with increasing $\tau_Q$). So large self-correlation time will inhabit the directional transport in $-x$ direction. When $\tau_Q$ is small($\tau_Q<0.5$), coupled passive particles($v_0=0.0$) are more easily produce directional transport then coupled self-propelled particles. But when $\tau_Q$ is large, coupled self-propelled particles are more easily produce directional transport then coupled passive particles.

\begin{figure}
\centering
\includegraphics[height=10cm,width=12cm]{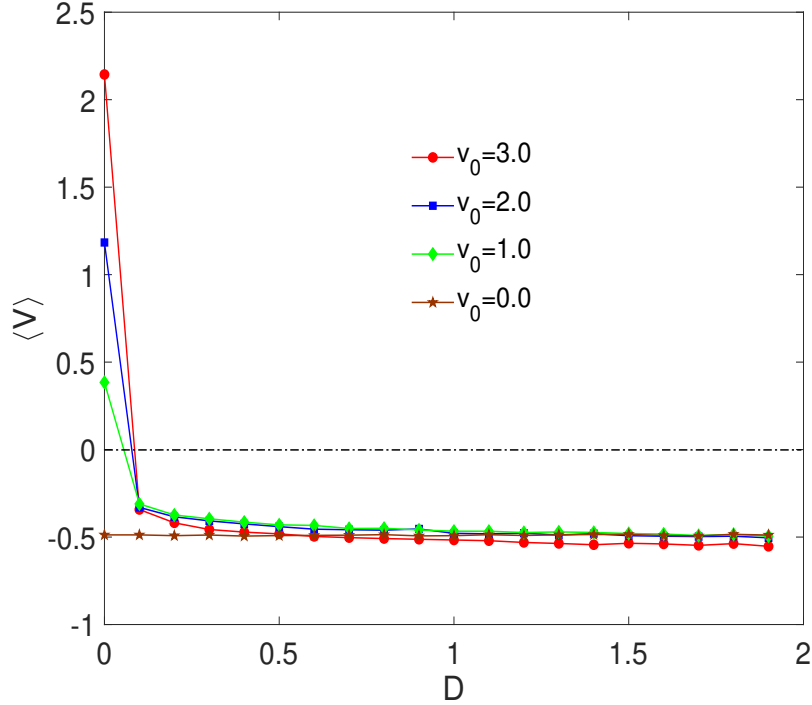}
\caption{The average velocity $\langle V\rangle$ as a function of the angle noise intensity $D$ with different $v_0$. The particles number $N=4$. The other parameters are $Q=1.0$, $\tau_Q=1.0$, $\tau_D=1.0$, $L=2.0$, $\Delta=0.4$, $k=0.5$, $a=1.0$.}
\label{VD}
\end{figure}

Fig.\ref{VD} shows the average velocity $\langle V\rangle$ as a function of the angle noise intensity $D$ with different $v_0$. In this figure, for passive particles($v_0=0$), the average velocity $\langle V\rangle<0$, so coupled passive particles move always in $-x$ direction, and angle noise intensity $D$ has negligible effect on these particles. For coupled propelled particles($v_0=1.0$, $v_0=2.0$ and $v_0=3.0$), we find $\langle V\rangle>0$ when $D=0$, but the moving direction changes to in $-x$ direction($\langle V\rangle<0$) with increasing $D$. So the coupled self-propelled particles move in $+x$ direction when the angle noise is not exit, and the moving direction changes from in $+x$ direction to in $-x$ direction with increasing angle noise intensity. This means the transport reverse phenomenon appears with increasing $D$ for coupled self-propelled particles. When $D>0$, we find the average velocity $\langle V\rangle$ decreases slowly with increasing $D$.

\begin{figure}
\centering
\includegraphics[height=10cm,width=12cm]{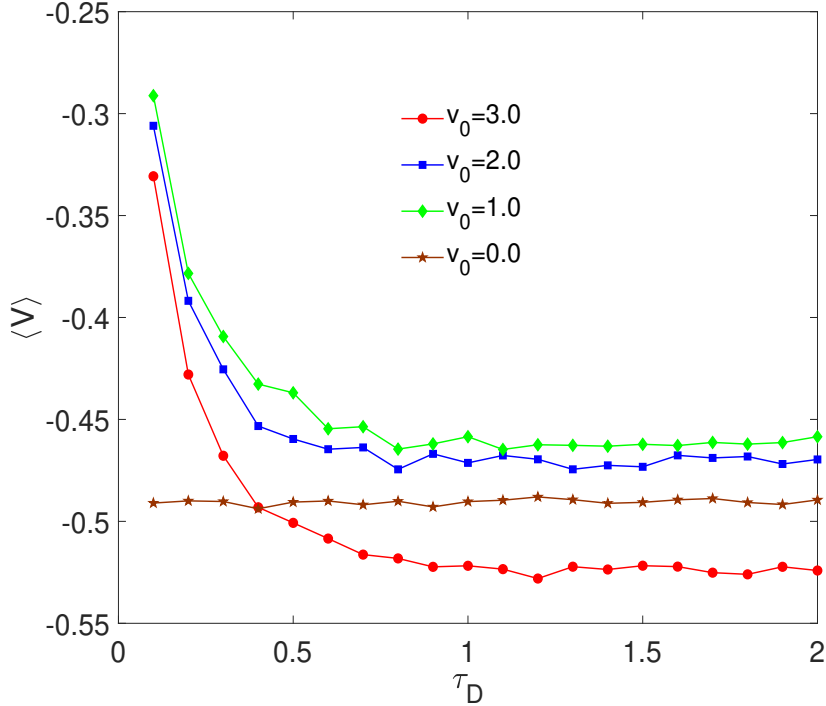}
\caption{The average velocity $\langle V\rangle$ as a function of the angle noise self-correlation time $\tau_D$ with different $v_0$. The particles number $N=4$. The other parameters are $Q=1.0$, $\tau_Q=1.0$, $D=1.0$, $L=2.0$, $\Delta=0.4$, $k=0.5$, $a=1.0$.}
\label{VtauD}
\end{figure}

Fig.\ref{VtauD} shows the average velocity $\langle V\rangle$ as a function of angle noise self-correlation time $\tau_D$ with different $v_0$. Just like the result of Fig.\ref{VD}, passive particles($v_0=0$) move in $-x$ direction, and $\langle V\rangle$ is almost remain unchanged with increasing $\tau_D$. But for propelled particles($v_0=1.0$, $v_0=2.0$ and $v_0=3.0$), we find $\langle V\rangle<0$, and $\langle V\rangle$ decreases with increasing $\tau_D$, and the slope of $\langle V\rangle-\tau_D$ changes to zero when $\tau_D$ is large.

\begin{figure}
\centering
\includegraphics[height=10cm,width=12cm]{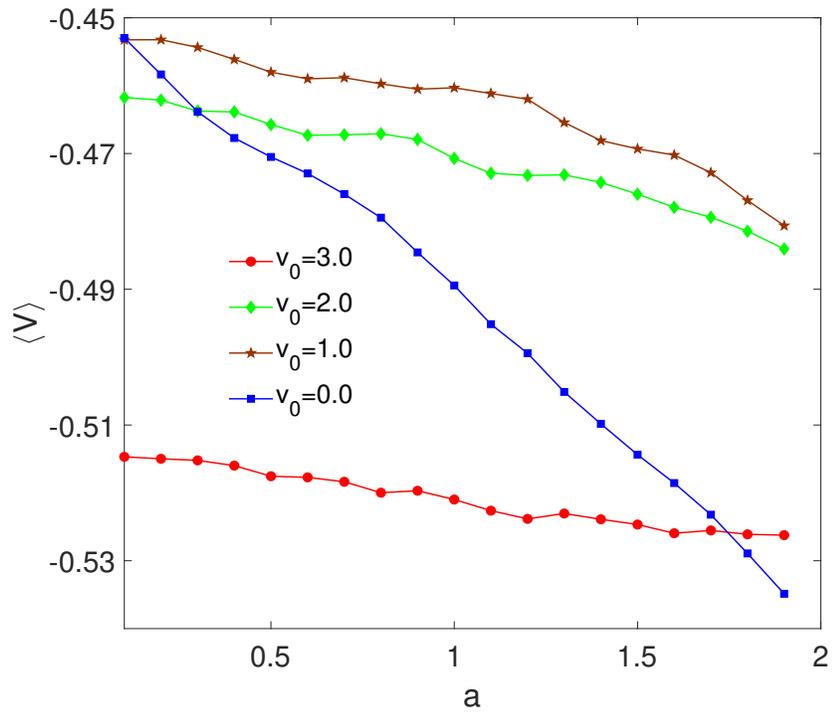}
\caption{The average velocity $\langle V\rangle$ as a function of the natural length $a$ of the spring with different $v_0$. The particles number $N=4$. The other parameters are $Q=1.0$, $\tau_Q=1.0$, $D=1.0$, $\tau_D=1.0$, $L=2.0$, $\Delta=0.4$, $k=0.5$.}
\label{Vxiaoa}
\end{figure}

The average velocity $\langle V\rangle$ as a function of the natural length of the springs $a$ with different $v_0$ is reported in Fig.\ref{Vxiaoa}. We find $\langle V\rangle<0$, and this result is compatible with the results of Figs.(\ref{VQ}, \ref{VtauQ}, \ref{VD}, \ref{VtauD}).  The average velocity $\langle V\rangle$ decreases with with increasing $a$. This is an interesting phenomenon, the longer of spring is, maybe the larger of the moving speed. We also find the absolute value of the slope for passive particles($v_0=0.0$) is larger than propelled particles($v_0=1.0$,$v_0=2.0$,$v_0=3.0$). So directional moving speed of coupled passive particles is more easily effect by the springs length than coupled self-propelled particles.

\begin{figure}
\centering
\includegraphics[height=10cm,width=12cm]{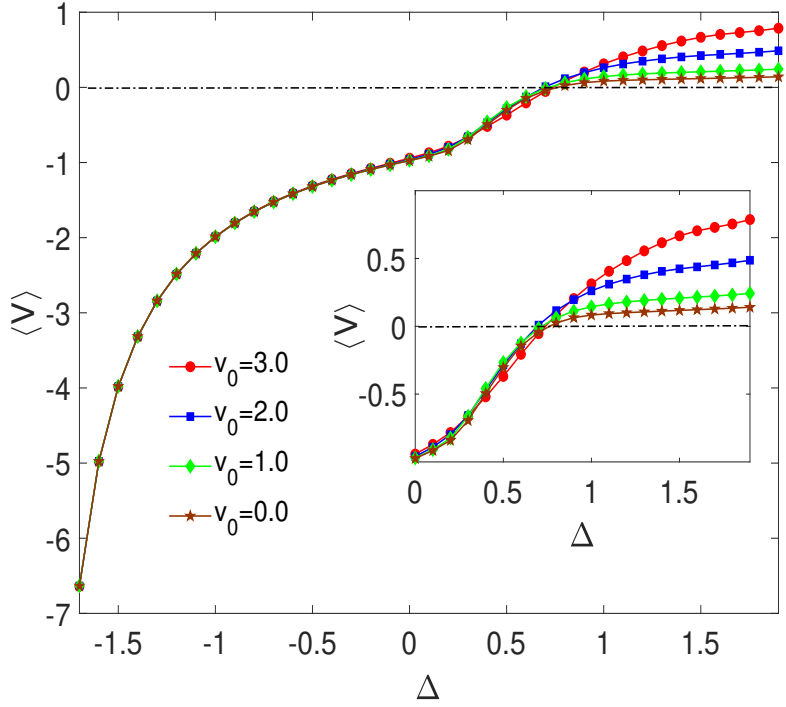}
\caption{The average velocity $\langle V\rangle$ as a function of the asymmetry parameter $\Delta$ with different $v_0$. The particles number $N=4$. The other parameters are $Q=1.0$, $\tau_Q=1.0$, $D=1.0$, $\tau_D=1.0$, $L=2.0$, $k=0.5$, $a=1.0$.}
\label{VDelta}
\end{figure}

Fig.\ref{VDelta} shows $\langle V\rangle$ as a function of asymmetry parameter $\Delta$ with different $v_0$. We find $\langle V\rangle<0$ when $\Delta<0.7$, and this result agrees well with the results of Figs.\ref{VQ}-\ref{VDelta} as $\Delta=0.4$ in those figures. The $\langle V\rangle-\Delta$ curves for different $v_0$ are almost coincide when $\Delta<0.7$. So the effect of $v_0$ will weak when $\Delta<0.7$. We can also find $\langle V\rangle>0$ when $\Delta>0.7$, and $\langle V\rangle-\Delta$ curves separate with other for different $v_0$. So the transport reverse phenomenon appears with increasing $\Delta$, and effect of $v_0$ becomes obvious when $\Delta$ is large.
\begin{figure}
\centering
\subfigure{
\includegraphics[height=10cm,width=12cm]{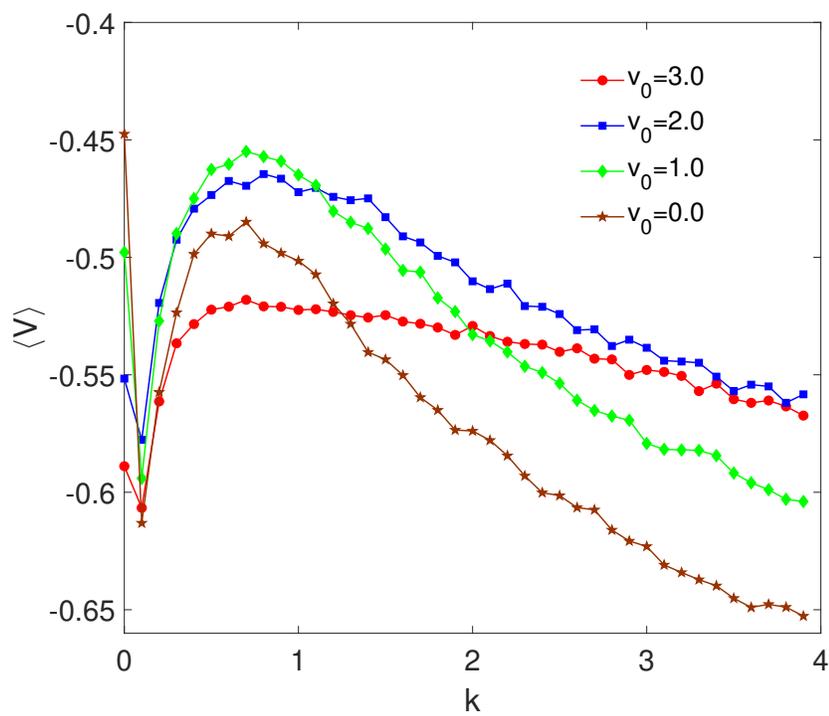}
}
\caption{$\langle V\rangle$ as a function of $v_0$ with different spring constant $k$. The particles number $N=4$. The other parameters are $Q=1.0$, $\tau_Q=1.0$, $D=1.0$, $\tau_D=1.0$, $\Delta=0.4$, $L=2.0$, $a=1.0$.}
\label{Vk2023}
\end{figure}

Fig. \ref{Vk2023} shows $\langle V\rangle$ as a function of the spring constant $k$ with different self-propelled speed $v_0$. We find $\langle V\rangle$ appears complex phenomenon with increasing $k$. $\langle V\rangle$ has a minimum and a maximum with increasing $k$. Proper small $k$ is good for the directional transport($k=0.1$), but the directional transport will be inhibited($|\langle V\rangle|$ has a minimum) when $k\approx0.7$, and then increasing $k$ promotes the directional transport($|\langle V\rangle|$ increases with increasing $k$ when $k>0.8$).

\begin{figure}
\center{
\includegraphics[height=10cm,width=12cm]{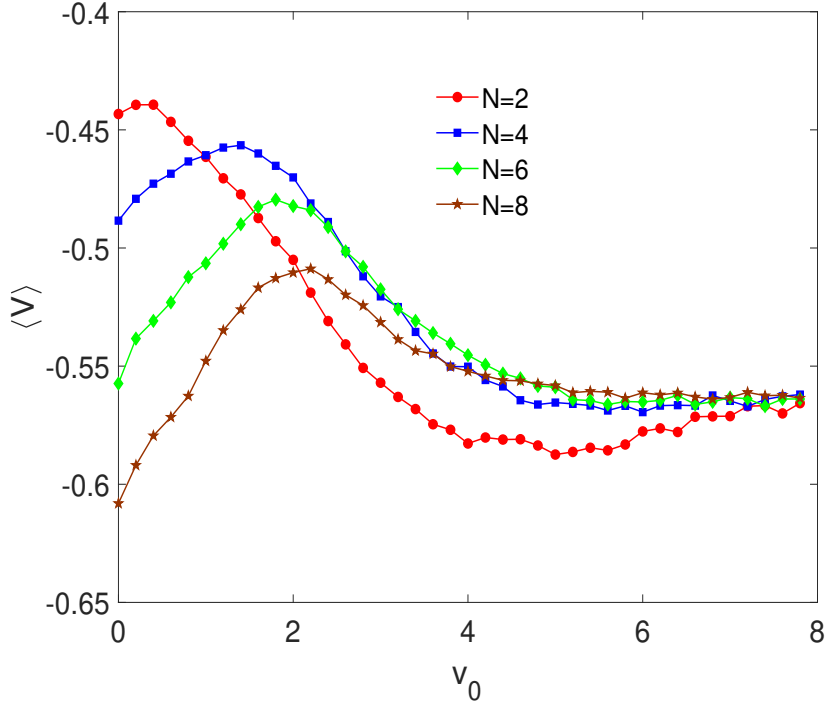}
\caption{The average velocity $\langle V\rangle$ as a function of self-propelled speed $v_0$ with different particle number $N$.  The other parameters are $Q=1.0$, $\tau_Q=1.0$, $D=1.0$, $\tau_D=1.0$, $L=2.0$, $k=0.5$, $a=1.0$.}
\label{Vv0}
}
\end{figure}
The dependence of $\langle V\rangle$ on the self-propelled speed $v_0$ with different coupling particle number $N$ is shown in Fig. \ref{Vv0}. We find the average $\langle V\rangle$ has a maximum($\langle V\rangle<0$) with increasing $v_0$ for different particle number($|\langle V\rangle|$ has a minimum $|\langle V\rangle|_{min}$ with increasing $v_0$). The more particles, the larger of $|\langle V\rangle|_{min}$ is. For passive particle($v_0=0$), the average speed $|\langle V\rangle|$ of $8$ coupled particles is larger then the average speed $|\langle V\rangle|$ of $2$($4$, $6$) coupled particles. So coupling is good for directional transport of passive particles. In this figure we can also find $|\langle V\rangle|\rightarrow-0.56$ as self-propelled speed $v_0\rightarrow 8$ whenever $N=2$, $N=4$, $N=6$ and $N=8$, so the effect of coupling  will become weak when $v_0$ is large.

\section{\label{label4}Conclusions}
In this paper, we numerically investigated the transport phenomenon of coupled self-propelled particles in the presence of potential with colored noise. We find large noise intensity of translational motion is good for the moving in $-x$ direction, but large self-correlation time of translational motion will inhibit the directional movement. Coupled passive particles move always in $-x$ direction when the asymmetry parameter $\Delta=0.4$, but coupled self-propelled particles changes the moving direction from in $x$ direction to in $-x$ direction with increasing angle noise intensity. Long spring is good for the directional movement of the particles. Whenever passive or self-propelled particles, the moving changes form in $-x$ direction to $+x$ direction with increasing asymmetry parameter. The average velocity appears complex behaviour with increasing spring constant $k$. For passive or very small self-propelled speed coupled particles, coupling is good for directional movement. For different coupling number, the average velocity tending to the same value when the self-propelled speed is large.

\section{Acknowledgments}

Project supported by Natural Science Foundation of Anhui Province(Grant No:1408085QA11) and College Physics Teaching Team of Anhui Province(Grant No:2019jxtd046).

\bibliographystyle{elsarticle-num}

\end{document}